\documentclass{aa}  

\usepackage{tabularx}
\usepackage{graphicx}
\usepackage{color}
\usepackage{xcolor}
\usepackage{txfonts}
\usepackage{mathrsfs}
\usepackage{listings}
\usepackage{hyperref}
\usepackage{amsmath}
\usepackage{url}
\usepackage{xspace}
\usepackage{soul}  
\usepackage{placeins}

\usepackage{glossaries}
\setacronymstyle{long-short}
\glsdisablehyper

\usepackage{CJKutf8}
\newcommand{\JO}{{\begin{CJK}{UTF8}{gbsn}(王加冕)\end{CJK}}}

\newacronym{MESA}{MESA}{Modules for Experiments in Stellar Astrophysics}

\newacronym{PDF}{PDF}{probability density function}

\newacronym{PCA}{PCA}{principal component analysis}
\newcommand{\PCA}{\gls{PCA}\xspace}

\newacronym{TESS}{TESS}{Transit Exoplanet Survey Satellite}

\newacronym{SPOC}{SPOC}{Science Processing Operations Center}

\newacronym{TIC}{TIC}{TESS Input Catalog}

\newacronym{HR}{HR}{Hertzsprung-Russell}

\newacronym{FP}{FP}{false positive}

\newacronym{TP}{TP}{true positive}

\newacronym{PE}{PE}{power excess}

\newacronym{RP}{RP}{repeating pattern}

\newacronym{PSD}{PSD}{power spectral density}
\newcommand{\PSD}{\gls{PSD}\xspace}

\newacronym{LC}{LC}{long cadence}

\newacronym{SC}{SC}{short cadence}

\newacronym{ACF}{ACF}{autocorrelation function}

\newacronym{ATL}{ATL}{Asteroseismic Target List}

\newacronym{RG}{RG}{red giant}

\newacronym{MS}{MS}{main-sequence}

\newacronym{SG}{SG}{subgiant}

\newacronym{FFI}{FFI}{full-frame image}

\newacronym{DR}{DR}{dimensionality reduction}
\newcommand{\DR}{\gls{DR}\xspace}

\newacronym{KDE}{KDE}{kernel density estimate}
\newcommand{\kde}{\gls{KDE}\xspace}

\newcommand{\muHz}{\,\mu\mathrm{Hz}}

\newcommand{\numax}{\nu_{\mathrm{max}}\xspace}
\newcommand{\dnu}{\Delta\nu\xspace}

\newcommand{\dotwo}{\delta\nu_{02}}

\newcommand{\Teff}{\,T_\mathrm{eff}}
\newcommand{\eps}{\,\varepsilon}
\newcommand{\bprp}{\,G_{\mathrm{BP}}-G_{\mathrm{RP}}}

\newcommand{\kepler}{\textit{Kepler}\xspace}

\newcommand{\pbjam}{\texttt{PBjam}\xspace}

\begin{document} 

   \title{Simplifying asteroseismic analysis of solar-like oscillators}
   \subtitle{An application of principal component analysis for dimensionality reduction}

   \author{M. B. Nielsen\inst{1}
          \and
          G. R. Davies\inst{1}
          \and
          W. J. Chaplin\inst{1}
          \and
          W. H Ball\inst{1,2}
          \and
          J. M. J. Ong \JO\inst{3,4}
          \and
          E. Hatt\inst{1}
          \and
          B. P. Jones\inst{1}
          \and
          M. Logue\inst{1}
          }

   \institute{School of Physics and Astronomy, University of Birmingham, Birmingham B15 2TT, UK\\
              \email{m.b.nielsen.1@bham.ac.uk}
              \and
              Advanced Research Computing, University of Birmingham, Birmingham B15 2TT, UK
              \and
              Institute for Astronomy, University of Hawai`i, 2680 Woodlawn Drive, Honolulu, HI 96822, USA
              \and
              Hubble Fellow
              }

   \date{Accepted 11/06/2023}

  \abstract
   {The asteroseismic analysis of stellar power density spectra is often computationally expensive. The models used in the analysis may require several dozen parameters to accurately describe features in the spectra caused by the oscillation modes and surface granulation. Many of these parameters are often highly correlated, making the parameter space difficult to quickly and accurately sample. They are, however, all dependent on a much smaller set of parameters, namely the fundamental stellar properties.}
   {We aim to leverage this to develop a method for simplifying the process of sampling the model parameter space for the asteroseismic analysis of solar-like oscillators, with an emphasis on mode identification.}
   {Using a large set of previous observations, we applied principal component analysis to the sample covariance matrix to select a new basis on which to sample the model parameters. Selecting the subset of basis vectors that explains the majority of the sample variance, we then redefined the model parameter prior probability density distributions in terms of a smaller set of latent parameters.}
   {We are able to reduce the dimensionality of the sampled parameter space by a factor of two to three. The number of latent parameters needed to accurately model the stellar oscillation spectra cannot be determined exactly but is likely only between four and six. Using two latent parameters, the method is able to produce models that describe the bulk features of the oscillation spectrum, while including more latent parameters allows for a frequency precision better than $\approx10\%$ of the small frequency separation for a given target.} 
   {We find that sampling a lower-rank latent parameter space still allows for accurate mode identification and parameter estimation on solar-like oscillators over a wide range of evolutionary stages. This allows for the potential to increase the complexity of spectrum models without a corresponding increase in computational expense.}


   \maketitle

\section{Introduction}
Asteroseismology is the process of measuring the oscillation frequencies of stars. This is typically done by first observing the flux or radial velocity variations due to the oscillations on the stellar surface, and then analyzing the \PSD of the time series. The properties of the spectrum, including the oscillation frequencies, allow for quantities such as the stellar mass, radius, and age to be precisely measured \citep[see, e.g.,][]{SilvaAguirre2017, Cunha2021}. Solar-like oscillators in particular are numerous and ubiquitous along the lower main sequence and during the red-giant evolutionary phases \citep{Chaplin2014, Yu2018, Hon2019, Hatt2023}. The oscillations therefore yield constraints on a wide range of physics in the stars themselves, but also in any orbiting companions and the environments they are embedded in, such as stellar clusters and associations.

Solar-like oscillators pulsate due to standing waves that permeate the stellar interior \citep[see, e.g.,][]{Chaplin2013}. The restoring force for these oscillations is the gradient of pressure, and so these standing waves are often simply called p modes. Many modes can be excited simultaneously and thus become visible on the stellar surface, and each one is distinguished by its spherical harmonic angular degree, $l,$ and azimuthal order, $m$, as well as a radial order, $n$. The most readily visible asteroseismic quantities are the regular spacing, $\dnu$, of the mode peaks and the frequency of maximum power, $\numax$. Measuring these global asteroseismic properties is computationally inexpensive \citep[see, e.g.,][]{Chontos2021, Hatt2023} but is sufficient for determining the stellar mass and radius to a precision of a few percent \citep{Chaplin2014}. These features of the oscillation spectrum have been shown to strongly depend on the bulk stellar properties, such as the surface gravity and effective temperature \citep{Kjeldsen1995}, with additional variance potentially due to metallicity \citep{Viani2017, Li2022}.   

The precision on the inferred stellar properties is increased by measuring the individual mode frequencies \citep[see, e.g.,][]{Lebreton2014}. This involves first identifying which oscillations are present in the \PSD by assigning an angular degree and radial order of the mode. This can then be followed by evaluating the likelihood of a detailed spectrum model that includes information about the mode amplitudes and lifetimes in addition to the precise mode frequencies. For long, high-cadence time series such as those from the CoRoT \citep{Baglin2009}, \kepler \citep{Borucki2010} and Transit Exoplanet Survey Satellite \citep[TESS;][]{Ricker2015} missions, this process may become computationally expensive. Part of this computational expense comes from the quantity of the data, but a non-negligible expense also comes from the overall shape and volume of the posterior probability density function. Depending on the required precision and quantities of interest, the spectrum models may include anywhere from $\sim10-100$ parameters \citep[see, e.g.,][]{Appourchaux2012, Davies2016a, Lund2017, Benomar2018a, Hall2021}. These parameters all depend on the same fundamental stellar properties, and many of them are highly correlated. High-dimensional, highly correlated posterior distributions are often difficult to map, thereby adding to the computational expense of measuring the oscillation parameters. 

Problems involving parameterized models with many correlated parameters are amenable to \DR methods. These methods seek to identify a lower-rank, latent parameter space that may be more directly related to an underlying set of physical parameters. For asteroseismology, the application of \DR methods can simplify the process of mode identification by mapping the smaller, latent parameter space, rather than a model parameter space that consists of dozens of parameters. Several methods exist for performing \DR, such as feature selection \citep{Guyon2003}, autoencoders \citep{Goodfellow2016, Kingma2013}, and matrix factorization \citep{Lee1999}. 
Here we demonstrate the use of \PCA for \DR to simplify the process of performing mode identification and extracting information from the oscillation spectrum of solar-like oscillators. Briefly, given a sample of observational measurements of the spectrum model parameters, the \PCA method decomposes the sample covariance matrix into a set of eigenvectors with associated eigenvalues. Each eigenvector explains a fraction of the total variance in the sample, and those that explain the majority of the variance are assumed to predominantly represent real correlations between parameters. The remaining eigenvectors, which explain the least amount of variance, are likely due to noise in the measurements and are discarded. This yields a reduced parameter space from which samples can be drawn, which can be projected back into the full parameter space to subsequently evaluate the model likelihood given an observed \PSD spectrum. We leverage this to simplify the process of mode identification by using a generative spectrum model, where the parameters are known to correlate with the stellar fundamental parameters.

In Sect.~\ref{sec:pb} we briefly summarize the spectrum model, which is based on the \pbjam model used by \citet{Nielsen2021} but extended to include additional background noise terms. In Sect.~\ref{sec:pca} we describe how we use \PCA to perform \DR and its application to mode identification in asteroseismology. The performance of the method is shown in Sect.~\ref{sec:performance}, followed by concluding remarks in Sect.~\ref{sec:conclusions}.
 
\section{The full spectrum model} \label{sec:pb}
 \begin{figure*} 
    \centering
    \includegraphics[width=1\linewidth]{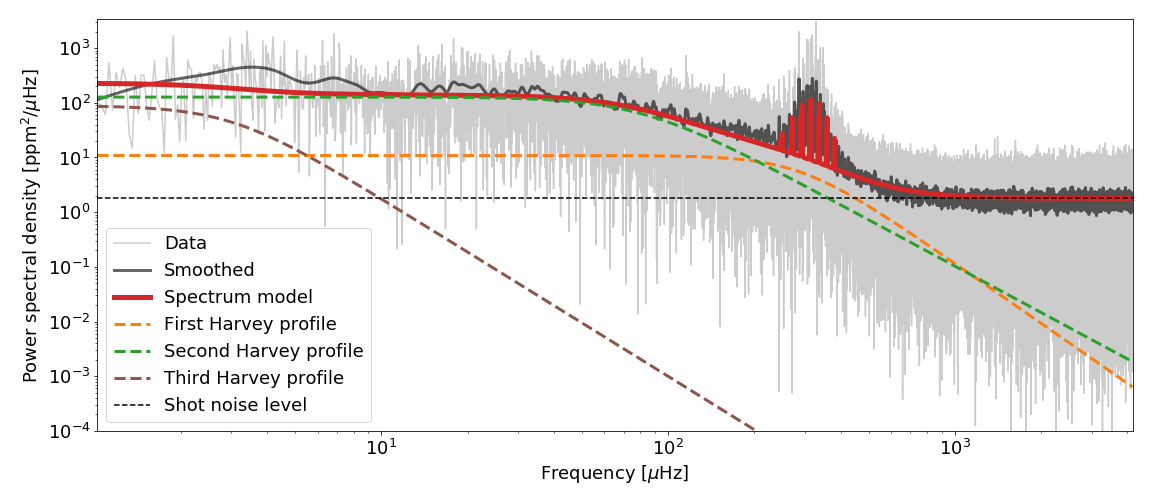}
    \caption{Example spectrum of the red giant $\epsilon$ Reticuli. The \PSD is shown in light and dark gray (smoothed). The components of the background model are shown with dashed lines, while the combined spectrum model, including the mode components, is shown in red. The parameters of the spectrum model are drawn from the prior sample around $\numax\approx300\muHz$, the equivalent to that of $\epsilon$ Reticuli.}
    \label{fig:example_spectrum}
\end{figure*}
Modeling the stellar oscillation spectrum is typically performed by drawing samples from the posterior distribution,
\begin{equation}
    P(\Theta|S) \propto \mathcal{P}(\Theta) \mathcal{L}(S|\Theta),
    \label{eq:posterior}
\end{equation}
where $\mathcal{L}(S|\Theta)$ is the likelihood of observing the \PSD spectrum, $S$, given a spectrum model that is a function of the set of parameters $\Theta$, and $\mathcal{P}(\Theta)$ is the prior representing our current knowledge of these parameters. By sampling the parameters of a generative spectrum model where the modes appear with a known pattern, we can evaluate the probability that an observed peak in the spectrum is due to a mode from the model and thereby assign a label of $n$ and $l$.

The parameter space is typically sampled using an Markov chain Monte Carlo approach \citep[e.g.,][]{Handberg2011, Lund2017} or nested sampling methods \citep{Corsaro2020}. In the following we use nested sampling, as this has been shown to perform well with multi-modal posterior distributions, which, depending on the chosen model, may be the case in a \PSD spectrum with multiple resolved modes \citep{MultiNest}. We used the \texttt{Dynesty} \citep{Speagle2020} package to perform nested sampling \citep[][]{Skilling2004}. This is, however, not a strict requirement and the presented method for \DR is applicable to other types of sampling or optimization methods. 

Here we restricted ourselves to using a \PSD model equivalent to that used by \citet{Nielsen2021}, with additional terms to describe the background power density not directly attributable to the oscillations. The spectrum model is defined as  
\begin{align} 
M\left(\Theta,\nu\right) =& \sum\limits_{n=1}^{N_O}\frac{h_{n,0}}{1+\frac{4}{\Gamma^2}(\nu-\nu_{n,0})^2}  +\frac{h_{n-1,2}}{1+\frac{4}{\Gamma^2}(\nu-\nu_{n-1,2})^2} \, + \\
&\sum\limits_{i=1}^{N_B} \frac{a_i / b_{i}}{1 + \left(\nu/b_i\right)^{c_i}} + W.
\end{align}
\label{eq:model}
Here the first sum is over the visible number of radial orders $N_O$ and consists of Lorentzian profiles describing the power density of each oscillation mode. The modes are characterized by their radial order $n$ and angular degree $l$, and the sum consists of pairs of modes defined by ($n$, $l=0$) and ($n-1$, $l=2$). The frequencies are parameterized by the asymptotic relation \citep[see, e.g.,][]{Mosser2015}
\begin{equation}
\begin{array}{ll}
\nu_{n,0}  & =\left(n+\eps+\frac{\alpha}{2}\left(n-n_{\mathrm{max}}\right)^2\right)\dnu\\
\nu_{n-1,2} & =\nu_{n,0}-\dotwo,
\end{array}
\label{eq:asymptotic}
\end{equation}
where $\eps$ is a phase term, $n_{\mathrm{max}}=\numax/\dnu-\eps$, and $\alpha$ is the scale of the second order mode frequency variation. The frequencies of the $l=2$ modes are offset from the $l=0$ modes by the small frequency separation $\dotwo$, which is kept as a free parameter but is identical for all mode pairs. The mode heights are parameterized by a Gaussian, 
\begin{equation}
\begin{array}{ll}
    h_{n,0} & = H_{\mathrm{E}}\exp\left(-0.5\left(\nu-\numax\right)^2 / W^2_{\mathrm{E}}\right), \\
    h_{n-1,2} & = 0.7\,h_{n,0},
\end{array}
\end{equation}
with a height $H_{\mathrm{E}}$, a width $W_{\mathrm{E}}$, and is centered on $\numax$. The heights, $h_{n-1,2}$, of the quadrupole modes are approximated as a fraction of the $l=0$ mode height, determined by the relative mode visibility. Here this fraction was set to $0.7$, which is appropriate for \kepler observations \citep[][]{Handberg2011}. Finally, we used a single value of the mode width $\Gamma$ for all the observed modes. 

The $l=1$ modes were omitted from the spectrum model since it is difficult to establish a precise and comprehensive description of their mode frequencies for a wide range of evolutionary stages. In addition we also did not consider modes of $l=3$ since these are typically only visible in the brightest targets. Inclusion of $l=1$ and $l=3$ modes is left for future work.
  
The last two parts of Eq.~\ref{eq:model} consist of the frequency-independent white noise level, $W$, and a number, $N_B$, of frequency-dependent background noise terms. The background terms are described by Lorentzian-like profiles centered on $\nu=0$ \citep{Harvey1985, Kallinger2014}. Each term has an amplitude, $a_i$, a frequency, $b_i$, and an exponent, $c_i$, which governs the decrease in the noise power with frequency. We used three background terms: The first two are typically attributed to the appearance of granulation on the stellar surface, and are most apparent in the spectrum at a characteristic frequency comparable to $\numax$ and at $\approx0.3\numax$, respectively. The third and lowest frequency term captures the combination of variability caused by, for example, long-term instrumental effects and stellar activity, which tend to dominate the \PSD at frequencies $\lesssim 1 \muHz$. An example of the spectrum model for the red giant $\epsilon$ Reticuli is presented in Fig.~\ref{fig:example_spectrum}, where we show the individual components of the background model along with the combined spectrum model.

The inclusion of the background term was motivated by the necessity to accurately model the background noise that the oscillation modes are embedded in. In addition, just like the oscillation modes, the granulation on the stellar surface that causes the background noise stems from convective processes in the star. The background noise parameters are therefore correlated with those of the oscillation modes, and so indirectly constrain the mode parameters \citep[see, e.g.,][]{Kjeldsen2011, Chaplin2011a}. 

The power density is $\chi^2$-distributed with two degrees of freedom. Given the model, $M$, in Eq.~\ref{eq:model} we can then write the log-likelihood function as
\citep[see, e.g.,][]{Woodard1984, Duvall1986}
\begin{equation}
 \ln{\mathcal{L}\left(S | \theta, \phi \right)} = - \sum\limits_{j=1}^J {\left(\ln M\left( {\theta, \phi ,\nu _j } \right) + \frac{{S_j }}{{M\left( {\theta, \phi ,\nu _j } \right)}}\right)} + \ln{\mathcal{L}}_{\mathrm{obs}}\left(d|\theta\right),
 \label{eq:likelihood}
\end{equation}
where $S$ is the observed power density spectrum, and $J$ is the total number of frequency bins in the spectrum. As in \citet{Nielsen2021}, we added an additional likelihood term, $\ln{\mathcal{L}}_{\mathrm{obs}}\left(d|\theta\right)$, which is the likelihood due to additional observational constraints, $d$, that are not directly part of the spectrum model. In this case $\mathcal{L}_{\mathrm{obs}}\left(d|\theta\right)$ is the joint probability given by normal distributions centered on the observed $\Teff$ and de-reddened $\bprp$, with a standard deviation corresponding to the respective uncertainties. We use values from \citet{Evans2018} in the following analysis. 
 
For clarity in the following we have separated the set of parameters $\Theta = \{ \dnu, \numax, \varepsilon, \dotwo, \alpha, H_{\mathrm{E}}, W_{\mathrm{E}}, \Teff, \bprp, \Gamma, a_i, b_i, c_i, W\}$ into two subsets $\theta$ and $\phi$. The subset $\theta$ are those parameters in $\Theta$ that we expect are predominantly functions of the intrinsic stellar parameters, whereas those in $\phi$ are functions that are dominated by extrinsic variability such as instrumental noise. We only applied \DR to $\theta$ as these are the parameters that are expected to be most strongly correlated with the stellar properties. These parameters are: all the mode parameters, those of the two high frequency Harvey-like background noise terms, and $\Teff$ and $\bprp$. The parameters $\phi$ (see Table \ref{tab:priors}) are the photon shot noise and the lowest-frequency background noise term. While these are to some extent correlated with the physical stellar properties, they are also strongly affected by instrumental effects such as the choice of pixel mask, pixel sensitivity, and observation duration. We therefore left these parameters out of the \DR. This means that we have $16$ parameters that are amenable to \DR and four that are left as independent random variables in the sampling.

\section{Defining the model parameter priors} \label{sec:pca}
Given the log-likelihood in Eq.~\ref{eq:likelihood} we next needed to define the prior probability density $\mathcal{P}\left(\Theta\right)$. We present the functions used as priors for $\phi$ in Table~\ref{tab:priors}. For $a_3$ in the instrumental background term and the white noise level, $W$, we used a normal distribution in logarithmic power. These distributions are centered on the means $\mu_p$ and $\mu_W$, respectively, which are determined by the power in the first and last few frequency bins of the spectrum. For the characteristic frequency $b_3$, the prior is a normal distribution in $\log{b_3}$, centered on $1\muHz$ with a width of $0.15\,\mathrm{dex}$. We used a $\beta$ distribution with shape parameters $\alpha=1.2$ and $\beta=1.2$ for the exponent in the background term, where we set the location and scale parameters to one and three, respectively.   
\begin{table}
    \caption{Additional model parameter priors.}
    \centering
    \begin{tabular}{cc}
        Parameters &  Prior density\\
        \hline
        $\log{a_3}$ & $\mathcal{N}(\mu_p, 2)$ \\
        $\log{b_3}$ & $\mathcal{N}(0, 0.15)$ \\
        $c_3$ & $\beta(1.2, 1.2, 1, 3)$ \\
        $\log{W}$ & $\mathcal{N}(\mu_W, 2)$ \\
    \end{tabular}
    \tablefoot{The parameters not included in the \DR are assigned independent one-dimensional prior probability density functions. The shape parameters of the $\beta$ distribution are $\alpha=1.2$ and $\beta=1.2$  and the location and scale parameters are one and three, respectively.}
    \label{tab:priors}
\end{table}

Next we needed to determine the prior probability densities for the parameter $\theta$ that are included in the \DR. We started by using a large sample of measurements of $\theta$, which is based on that of \citet{Nielsen2021} and consists of the $16$ model parameters for $13\,443$ targets observed in $30$-minute and $58$-second cadences by \kepler and $2$-minute cadence by TESS. Our assumption here is that, using \PCA on the sample covariance matrix, it is possible to identify a set of latent variables, $\hat{\theta}$, which is likely of smaller dimension than $\theta$. The prior density is then defined in terms of $\hat{\theta}$ instead of $\theta$ by projecting the original prior sample into the latent parameter space, and then approximating that distribution by a \kde. 

\subsection{Computing the weighted covariance matrix}
Given an $N\times D$ matrix, $\boldsymbol{\theta}$, consisting of $N$ observations of the $D=16$ model parameters, we first computed the weighted covariance matrix,
\begin{equation}
    \mathbf{C} = \frac{\sum_{n=1}^N w_n}{\left(\sum_{n=1}^N  w_n\right)^2 - \sum_{n=1}^N w_n^2}\boldsymbol{\theta^{\prime}}^T\,\mathbf{W}\,\boldsymbol{\theta^{\prime}},
\end{equation}
where $\boldsymbol{\theta^{\prime}}$ is $\boldsymbol{\theta}$ translated and scaled to zero mean and unit standard deviation in each column. The matrix $\mathbf{W}$ is an $N \times N$ diagonal matrix with elements $w_n$, which are the weights on each individual observation of the model parameters. We did not consider the more general case of different sets of weights for each parameter and observation as this does not allow for a simple analytical method of determining the principal components \citep[see, e.g.,][]{Tamuz2005}.

Since \PCA minimizes the variance of a given sample along each of the principal components, the proposed method has two shortcomings that the row weights $w_n$ can be used to mitigate. First, \PCA is sensitive to outliers in the sample. This can be treated by assigning low weights relative to the remainder of the sample. We do not identify outliers here as the sample of observations has been previously vetted manually. Secondly, \PCA is less effective as a \DR method if parameters in the sample are nonlinearly correlated, which is the case for our sample of model parameters since it spans a wide range of evolutionary stages. For example, the dependence of $\eps$ in Eq.~\ref{eq:asymptotic} on $\Teff$ changes significantly between main-sequence and red-giant stars \citep[see, e.g.,][]{White2012}. 

\begin{figure}
    \centering
    \includegraphics[width=\linewidth]{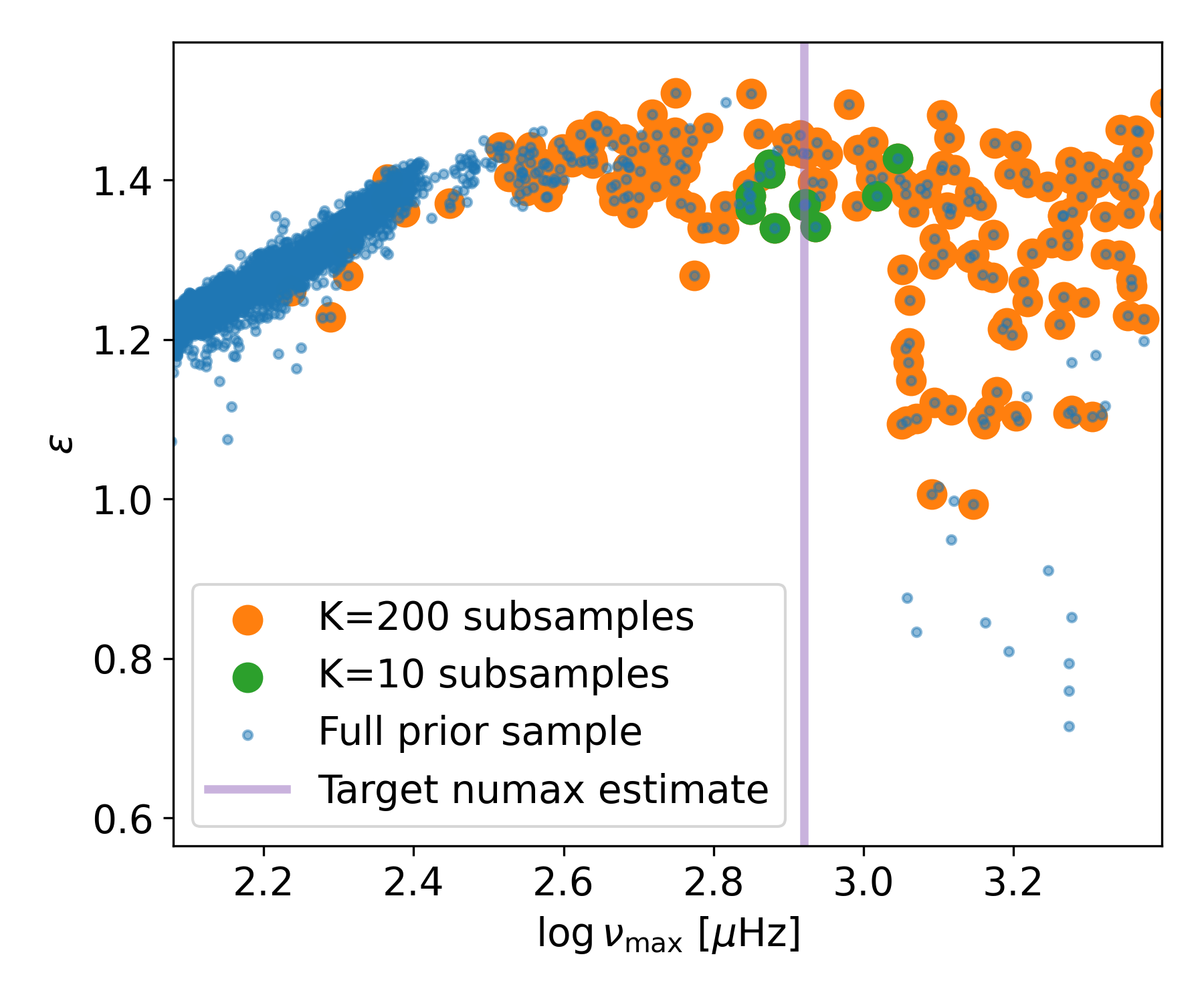}
    \caption{$\epsilon$ and $\log{\numax}$ values for a selection of the $K=10$ (green) and $K=200$ (orange) nearest neighbors, determined by the Euclidean distance relative to an example target with a notional $\numax\approx830\,\muHz$ (horizontal purple line), an effective temperature $\Teff\approx5640\,$K, and $\bprp\approx0.86$.}
    \label{fig:selection}
\end{figure}
 
However, these nonlinear effects in the parameter correlation are reduced when only considering targets that are similar to the target of interest. This is done by setting the row weights $w_n$ such that they are unity for targets in a small volume around the expected target parameters, and zero elsewhere. This small volume of targets is defined by using the $K$ nearest neighbors in terms of the Euclidean distance between the estimated target values of $\numax$, $\Teff$, and $\bprp$, and those of the prior sample. Figure \ref{fig:selection} illustrates the process of selecting subsamples around a notional target with $\numax\approx830\,\muHz$, $\Teff\approx5640\,$K, and $\bprp\approx0.86$. In this case, the correlation is strongly nonlinear on the $\epsilon$ and $\log{\numax}$ plane over large ranges of $\numax$, but for decreasing $K$ the correlation becomes predominantly linear.

We find that with the current sample of prior observations, the method is insensitive to the exact choice of $K$, which can be $\sim10-1000$. This selection scheme does require some prior knowledge of $\numax$ for the target. However, this can be estimated to within a few percent by for example fitting simple models to the \PSD \citep[e.g.,][]{Huber2011, Chontos2021}, or to within $\sim 10 \%$ by using the asteroseismic scaling relations \citep{Kjeldsen1995} along with an approximate stellar radius from parallax and broad-band photometry \citep[see, e.g.,][]{Campante2019, Hatt2023}. 

 \subsection{Sampling the latent parameter space}
 \begin{figure*}
    \centering
    \includegraphics[width=0.405\linewidth]{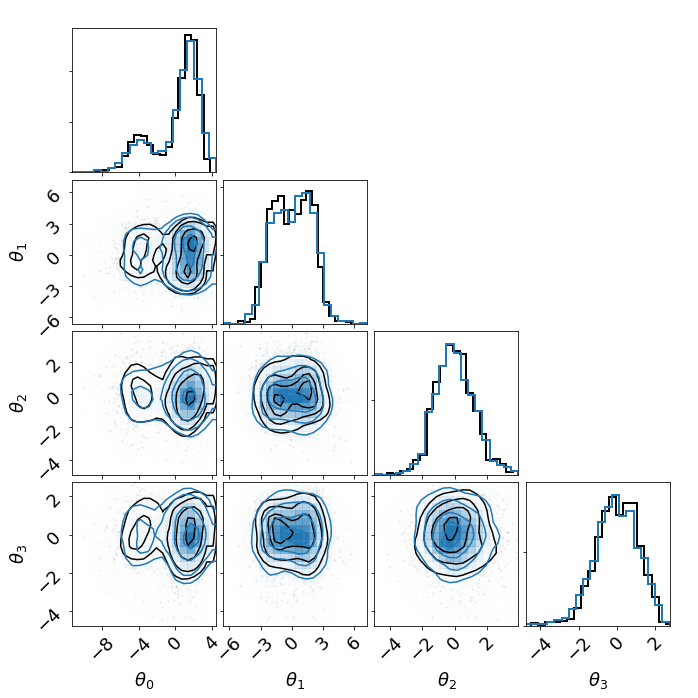}
    \includegraphics[width=0.59\linewidth]{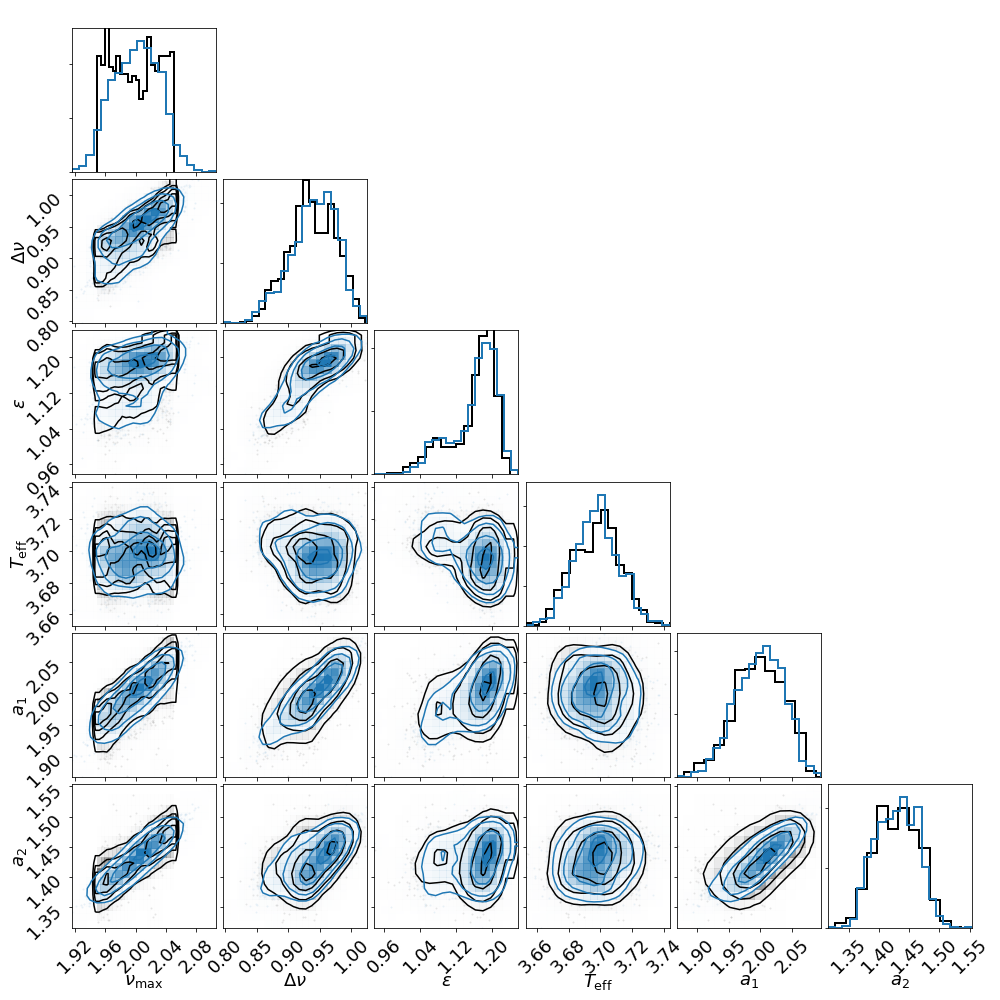}
    \caption{Corner plots of a sample of 400 targets around $\numax=100\muHz$ drawn from the prior distribution. Left: Projection of the sample into the latent parameter space (in black) and a corresponding sample (in blue) drawn from the one-dimensional \kde approximations of the marginalized distributions (diagonal frames). Right: Original 400 prior sample draws (in black) with the sample drawn from the latent space, projected into the model parameter space (in blue). For clarity we only show a subset of the model parameters, where all parameters except $\eps$ are in base-10 logarithmic units.}
    \label{fig:projectedPrior}
\end{figure*}
The principal components are found by identifying the eigenvectors, $\mathbf{Q}$, by factoring $\mathbf{C}$ as
\begin{equation}
\mathbf{C} = \mathbf{Q} \boldsymbol{\Lambda} \mathbf{Q}^{-1},
\end{equation}
where the columns of $\mathbf{Q}$ are the eigenvectors (principal components), and the diagonal matrix $\boldsymbol{\Lambda}$ consists of their eigenvalues. We used the \texttt{Numpy} library \citep{Harris2020} to perform the eigendecomposition. 

The set of eigenvectors, $\mathbf{Q}$, forms a new basis for the sample, where each axis corresponds to a latent variable. Any given set of observations $\theta_i^{\prime}$ can subsequently be projected onto the new basis by
\begin{equation}
    \hat{\theta}_i = \mathbf{Q}^T \theta_i^{\prime\,T}.
    \label{eq:transform}
\end{equation}
Conversely, a sample drawn from the eigenvector space can be projected back to the model parameter space by
\begin{equation}
    \theta_i^{\prime\,T} =  \hat{\theta}_i\, \mathbf{Q}^T.
    \label{eq:invTransform}
\end{equation}

Using Eq.~\ref{eq:transform}, we can transform draws from the prior sample in the model parameter space, into the latent parameter space to show their distribution on the new basis. Figure~\ref{fig:projectedPrior} shows 400 draws from around $\numax=100\muHz$ projected onto a subset of the new basis. We can use the distribution of these samples to approximate what the prior density looks like in the latent space, and subsequently start drawing new samples from this latent prior density. 

For simplicity, we only considered the marginalized distributions of the projected prior, where we used a one-dimensional \kde to approximate the distribution of each latent parameter. As seen by the marginal distribution of $\theta_2$, for example, this approach can lead to a small loss of accuracy in representing details such as multi-modality, which is caused by selecting a large \kde bandwidth. However, as shown in the right set of frames in Fig.~\ref{fig:projectedPrior} the original prior sample is still well represented by this sample when projected into the model parameter space using Eq.~\ref{eq:invTransform}. Sharp features such as discontinuities in the model parameter prior are however not well captured by this method, as indicated by the marginal distributions of $\numax$. This loss of fidelity around sharp features in the prior samples simply translates into a reduction in efficiency of the sampling, since more samples will be drawn from areas of parameter space that the prior sample suggests should be improbable.

\begin{figure}
    \centering
    \includegraphics[width=\linewidth]{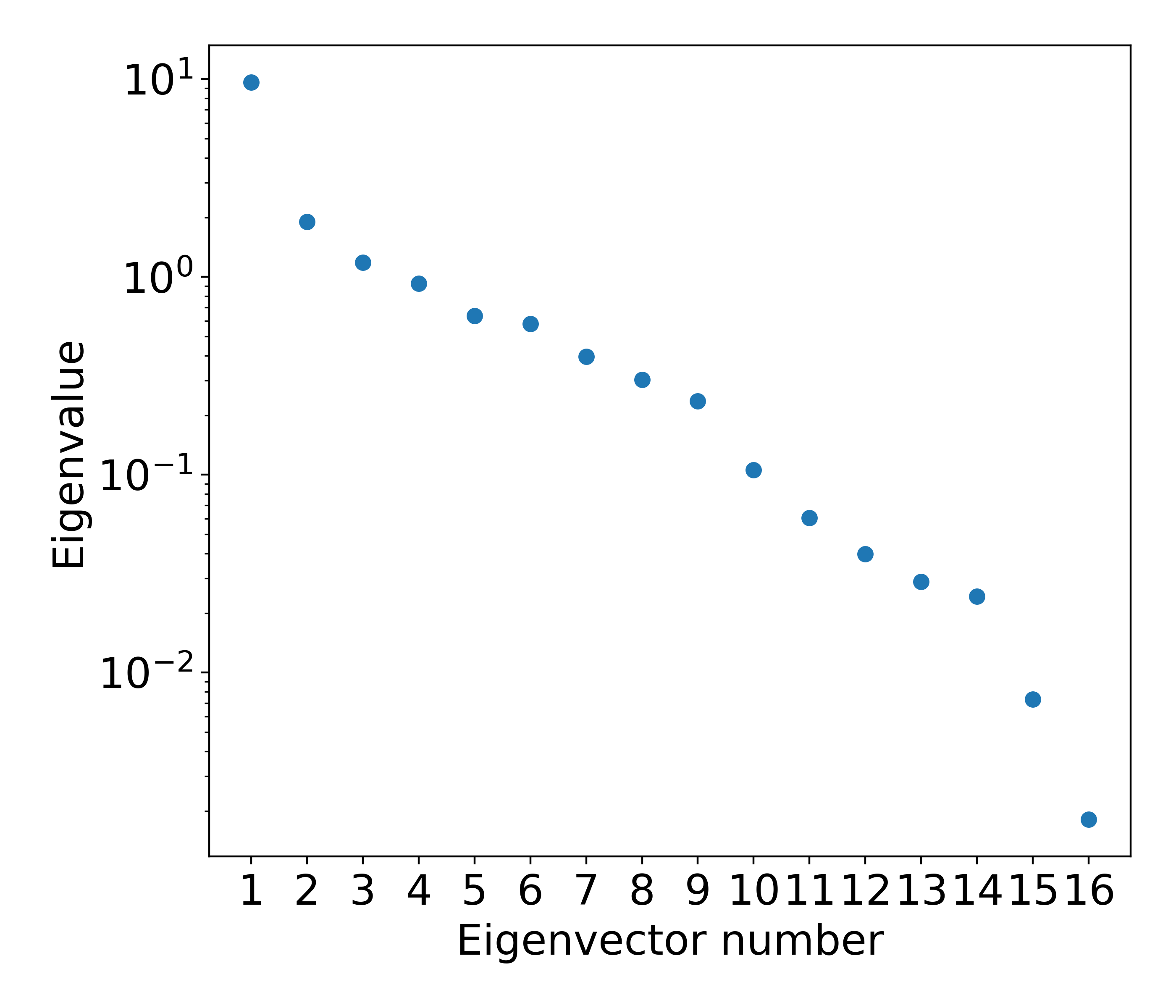}
    \caption{Eigenvalues of a covariance matrix computed from 400 samples of $\boldsymbol{\theta}$ around and input $\numax = 100\muHz$, where $\boldsymbol{\theta}$ consists of $16$ parameters. The eigenvalues are ordered by descending value, and the first few corresponding eigenvectors explain the majority of the sample variance.}
    \label{fig:skree}
\end{figure}

The eigenvalues determine the amount of the total variance in the sample that is explained by each corresponding eigenvector. The eigenvectors that explain the majority of the correlation can therefore be chosen by selecting the $d$ vectors with the highest eigenvalues, where the variance along the remaining eigenvectors is assumed to be due to noise. Figure~\ref{fig:skree} shows an example of the eigenvalues of $\mathbf{C}$ ordered according to value. In this case the majority of the variance is explained by the eigenvectors associated with the first few eigenvalues, while including additional eigenvectors accounts for comparatively less variance.  

\begin{figure}
    \centering
     \includegraphics[width=\linewidth]{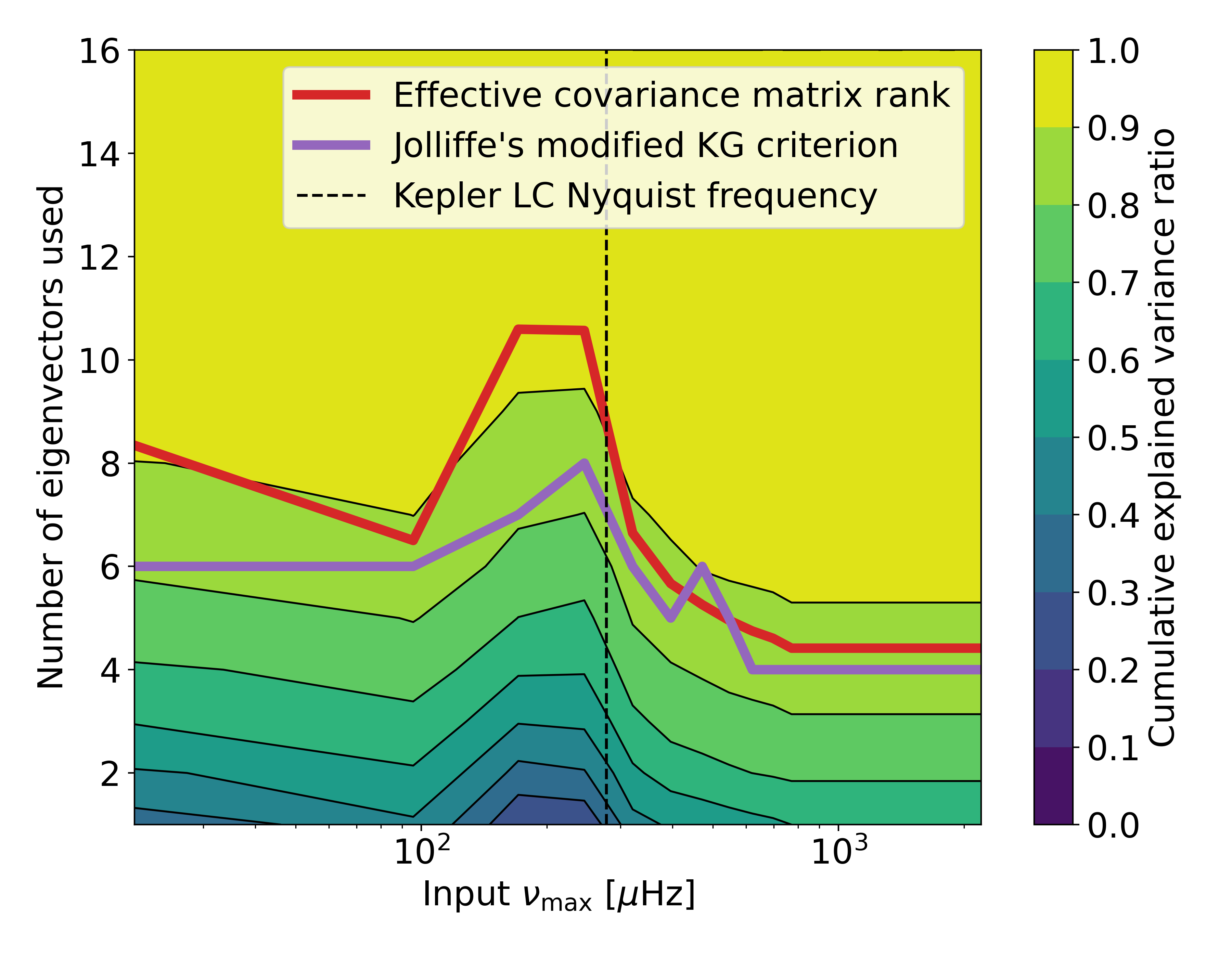}
    \caption{Cumulative explained variance of 400 samples of $\boldsymbol{\theta}$ for different combinations of input $\numax$ and selected eigenvectors. The dashed line indicates the delineation between stars observed in \kepler long cadence (low frequency) and short cadence (high frequency). The effective covariance matrix rank is indicated in red, and the number of latent parameters suggested by Jolliffe's modified KG criterion is shown in purple.}
    \label{fig:cumulativeskree}
\end{figure}

Picking a number, $d$, of eigenvectors where $d < D$, necessarily means discarding information from the sample corresponding to the missing variance along the remaining $D-d$ vectors. The amount of variance retained by a choice of $d$ is shown by the cumulative explained variance shown in Fig. ~\ref{fig:cumulativeskree}, where the variation in $\numax$ is due to the local density and covariance of the prior sample targets.

\subsection{Picking the number of latent parameters}
 From a data-driven perspective, it is possible to estimate the approximate value of $d$ with several different methods \citep[see, e.g.,][for a comparison of several methods]{Cangelosi2007}. Here we show the use of two such methods: Jolliffe's modification to the Kaiser-Guttman \citep[KG;][]{Guttman1954, Kaiser1960, Jolliffe1972} criterion and the effective matrix rank estimation.

The KG criterion suggests using eigenvectors of the correlation matrix \footnote{In our case, when each model parameter is normalized to unit variance, and with weights $w_n=1$, the covariance and correlation matrices are identical.} with eigenvalues greater than $1$, since these explain variance in the model space corresponding to more than one parameter. \citet{Jolliffe1972} later found that using eigenvalues greater than 0.7 is more appropriate based on simulated data sets, since this retains a greater fraction of the total variance and so potentially discarding less information.

An alternative to the KG criterion is the effective covariance matrix rank, $E$, which is related to the entropy of the eigenvalues \citep[][]{Roy2007} and is given by
\begin{equation}
    E = \exp{\left[-\sum_{d=1}^D \lambda_d \ln{\lambda_d}\right]},
\end{equation}
where $\lambda_d = \Lambda_d/ \sum_{d=1}^D \Lambda_d$, and the exponent is the Shannon entropy \citep{Shannon1948} of the eigenvalues. The effective rank is a non-integer value that approaches $1$ when the entropy of the eigenvalues is large, which corresponds to only a few eigenvalues explaining most of the variance. Conversely, the effective rank approaches $D$ when all the eigenvalues entropy is small, meaning they are all equally important.

In Fig.~\ref{fig:cumulativeskree} we compare the number of eigenvectors suggested by both methods with the cumulative explained variance. While these two methods suggest different numbers for the eigenvectors to retain, the typical value for our sample is between four and eight, which corresponds to explaining $80$--$90\%$ of the total sample variance, with the effective matrix rank method suggesting ten latent parameters is necessary between $\numax\approx100,\muHz$ and $200\,\muHz$.

However, neither the effective covariance matrix rank nor the modified KG criterion use any physical information about the model parameters involved. It is typically assumed, in the statistical modeling of the structure and evolution of solar-like oscillators, that the observed parameters generated by a star can be described using only a handful of fundamental parameters. By accepting a certain level of uncertainty, it is possible, for example, to describe a star and thereby the predicted mode frequencies using just its initial mass, initial metallicity, and current age. This assumes a degree of simplification, for example a known helium enrichment law, or other initial chemical relation in r-process elements. If we were to relax these assumptions to achieve a higher degree of accuracy, we then introduce additional parameters to describe the observed properties of the star. 

This process of relaxing assumptions and adding more parameters can in principle be continued to reach a greater degree of detail, at the cost of making the stellar model more complex. However, an analogous study with numerical models of stellar structure (which we describe in more detail in \autoref{app:model_var}) indicates that perhaps five or so free parameters are already more than sufficient to describe the effects on the mode frequencies of changing a larger number of parameters of these models, associated with quite different physical processes and properties. While the stellar model parameters are not the same as the latent parameters identified by the \PCA method, the precision achieved by stellar modeling with just five parameters suggests that the number of sufficient latent parameters is substantially smaller than the 16 model parameters, and possibly very similar.
  
\section{Performance of the PCA method} 
\label{sec:performance}
Our physical motivation suggests that the number of latent parameters should be small compared to the number of model parameters. However, any number of latent parameters less than the full rank of the covariance matrix potentially reduces the precision of the mode identification. In the following we investigate how the choice of the number of latent parameters influences the performance of the mode identification and estimation of the global parameters.
\begin{figure*}
    \centering
    \includegraphics[width=0.99\linewidth, trim={0 20 0 0}, clip]{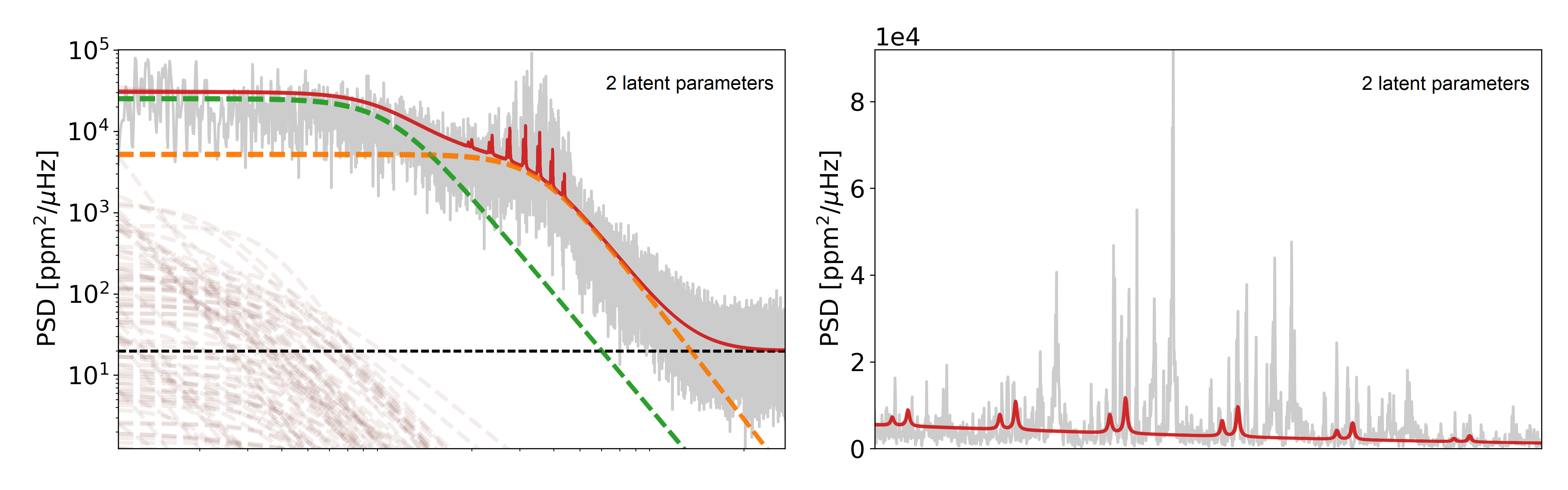}
    \includegraphics[width=0.99\linewidth, trim={0 20 0 0}, clip]{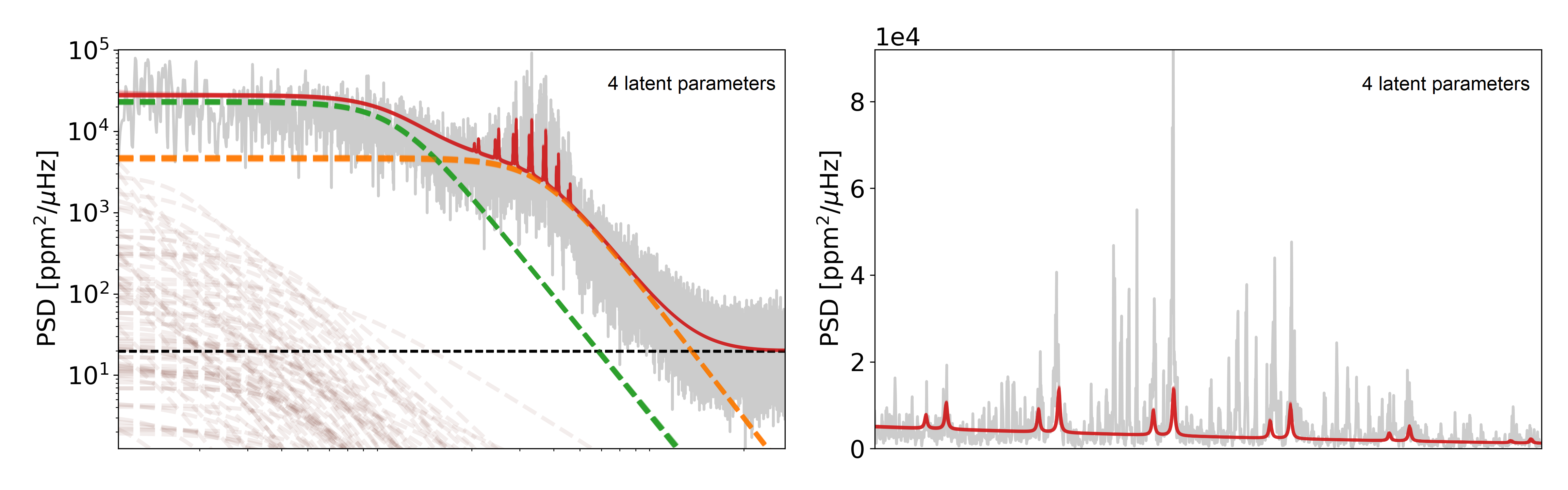}
    \includegraphics[width=0.99\linewidth, trim={0 20 0 0}, clip]{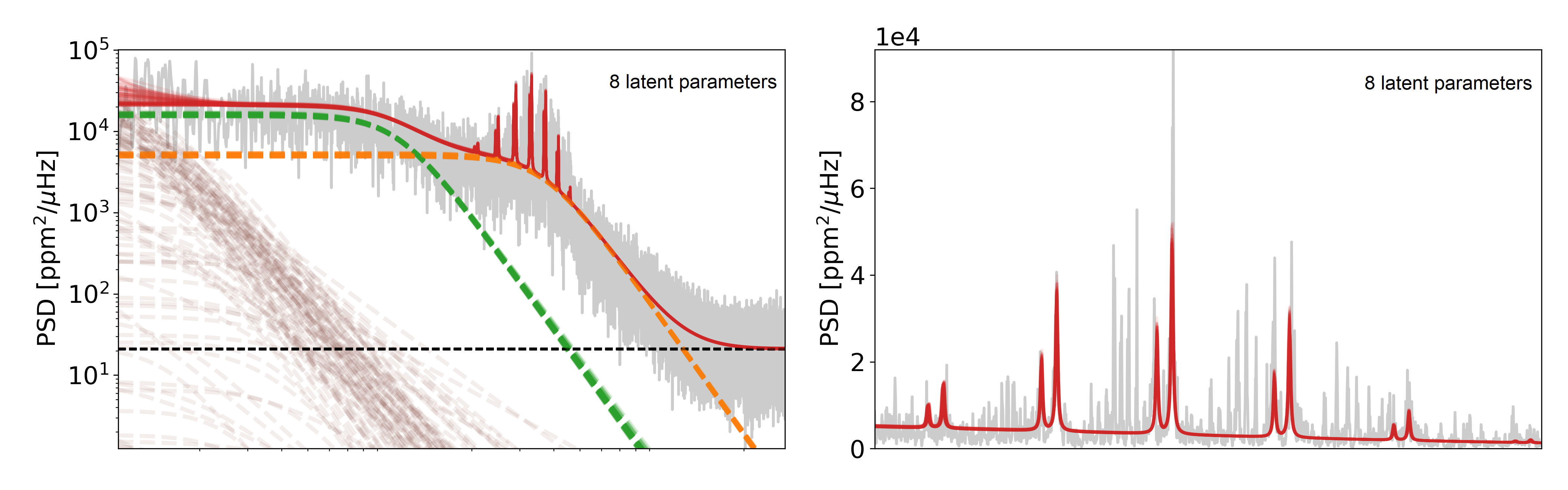}
    \includegraphics[width=0.99\linewidth]{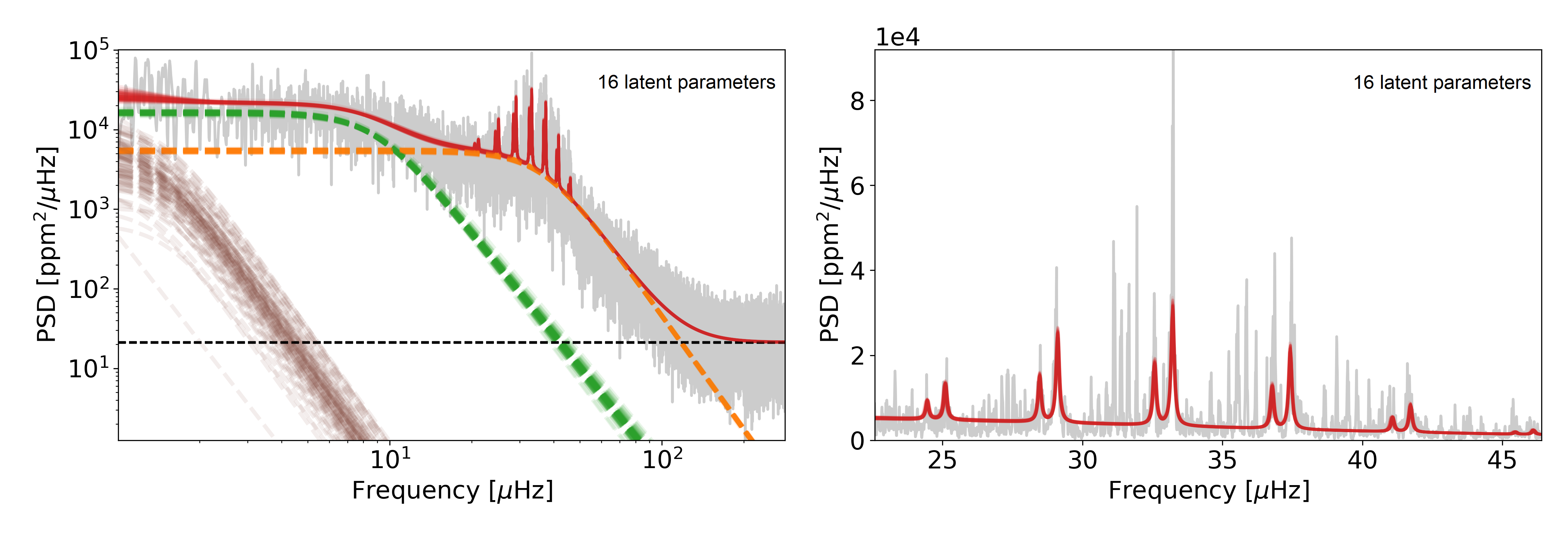}
    \caption{Examples of the smoothed spectrum of KIC6863017 (gray) compared to models that use parameters drawn from the posterior distribution with different numbers of latent parameters. Left column: Wide view of the spectrum, showing the background terms, which have the same color-coding as in Fig.~\ref{fig:example_spectrum}, with the combined model samples in red. Right column: Zoomed-in view of the p-mode envelope, showing the individual Lorentzian profiles for the modes.}
    \label{fig:example_results}
\end{figure*}

Figure~\ref{fig:example_results} shows an example of models generated from the posterior distribution of $\Theta$ for a typical red giant (KIC6863017) from our sample. Despite only using two latent parameters, along with the third Harvey profile and white noise components, the \PCA method is able to generate models that are broadly comparable to the observed spectrum, both in terms of the background noise and the p-mode envelope. We find that, in general, using only two latent parameters is sufficient to identify the bulk features in the spectrum, such as the first and second background terms, and the shape and location of the p-mode envelope. These features are strong functions of the surface gravity and effective temperature of the star, which in turn depend predominantly on the mass and evolutionary stage. Two latent parameters are, however, insufficient to accurately capture the covariance of parameters that govern the detailed frequency locations, such as $\dotwo$ and $\eps$. This is only achievable using $d>2$ latent parameters.

\begin{figure}
    \centering
    \includegraphics[width=\linewidth]{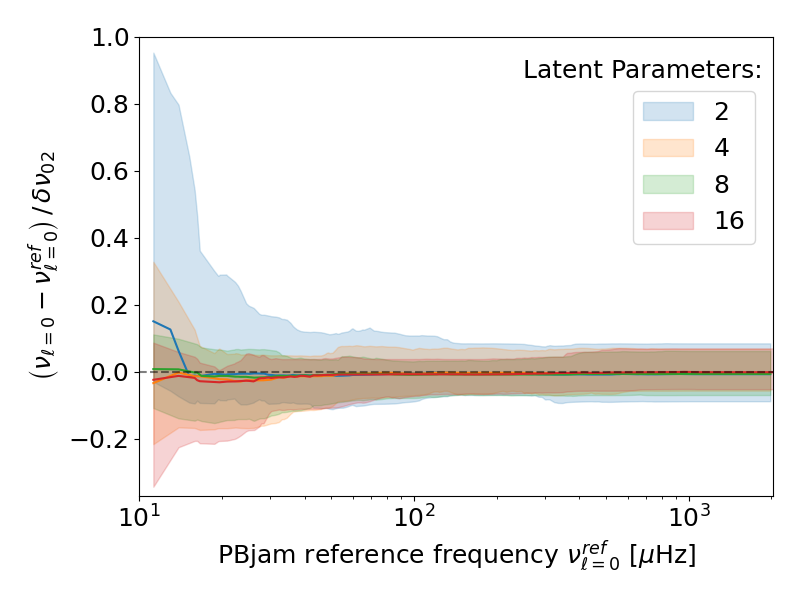}
    \caption{Mode identification precision for a sample of $50$ stars using different numbers of latent parameters. The differences between the radial mode frequencies of the \PCA method and \pbjam are shown in relation to that of $\dotwo$ from \pbjam. The shaded regions indicate the $68\%$ interval of the observed frequency differences for different numbers of latent parameters used in the model, and the solid colored lines show the median values in each frequency bin. For reference, the dashed line denotes complete agreement between the sets of mode frequencies.}
    \label{fig:freq_compare}
\end{figure}
Figure~\ref{fig:freq_compare} shows a comparison of the radial mode frequencies measured using the \PCA method and those from \pbjam for a sample of $50$ stars linearly spaced in $\log{\numax}$. The \pbjam solutions for these targets were manually vetted to ensure they agree well with the observed spectrum. The range of frequency differences are shown in relation to $\dotwo$ from \pbjam to indicate the quality of the \PCA mode identification. Radial modes that differ from the \pbjam solution by a large fraction of $\dotwo$ are more likely to be incorrect identifications. We used seven radial orders to compute the frequency differences. In some cases, the combination of $\numax$, $\dnu$, and $\eps$ leads to estimates of the radial orders $n$ that differ by $\pm 1$ compared to \pbjam. To account for this we computed the \pbjam solutions for nine radial orders and find the set of modes that best match the seven computed by the \PCA method. 

Using two latent parameters the \PCA method tends to overestimate the radial mode frequencies by a large fraction of $\dotwo$ for evolved red-giant stars with $\numax\approx20\muHz$. However, above $\sim100\muHz$ the frequency differences are $\approx10\%$ of $\dotwo$ regardless of the number of latent parameters used. We find that the discrepancy is decreased to $\approx5\%$ of $\dotwo$ when only the central five modes pairs are considered, and $\approx3\%$ of $\dotwo$ when the central three modes pairs are considered. 

\begin{figure}
    \centering
    \includegraphics[width=\linewidth, trim={0 30 30 65}, clip]{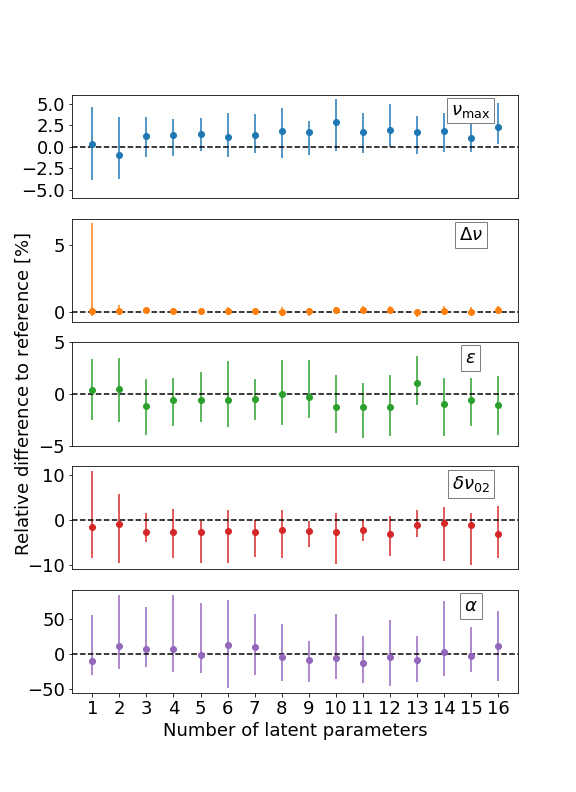}
    \caption{Relative differences of the asymptotic relation parameters determined by \pbjam and the \PCA method using different numbers of latent parameters. The dashed black lines denote agreement between the two methods. The error bars denote the spread in values for the same sample of $50$ stars used in Fig.~\ref{fig:freq_compare}.}
    \label{fig:params_compare}
\end{figure}
Figure~\ref{fig:params_compare} shows a comparison between the measured asymptotic relation parameters from the \PCA method and the \pbjam values for the same $50$ targets shown in Fig.~\ref{fig:freq_compare}. Generally, with more than two latent parameters there is little difference between the observed distributions. The observed parameters show a bias of $\approx1.5\%$ in $\numax$ and $\approx3\%$ in $\dotwo$ for more than two latent parameters. While this bias is absent when using only one or two latent parameters the scatter is typically larger, where the most extreme outlier is in $\dnu$. For two latent parameters the median difference in the observed $\dnu$ is $\approx 0.07\%$. The large differences in the observed values of $\alpha$ are consistent with the modes furthest from $\numax$ showing the largest differences compared to the \pbjam reference frequencies. However, we do not expect an exact match between the method presented here and \pbjam since the two methods differ in the way they encode the prior information and subsequently perform the sampling.

\begin{figure}
    \centering
    \includegraphics[width=\linewidth]{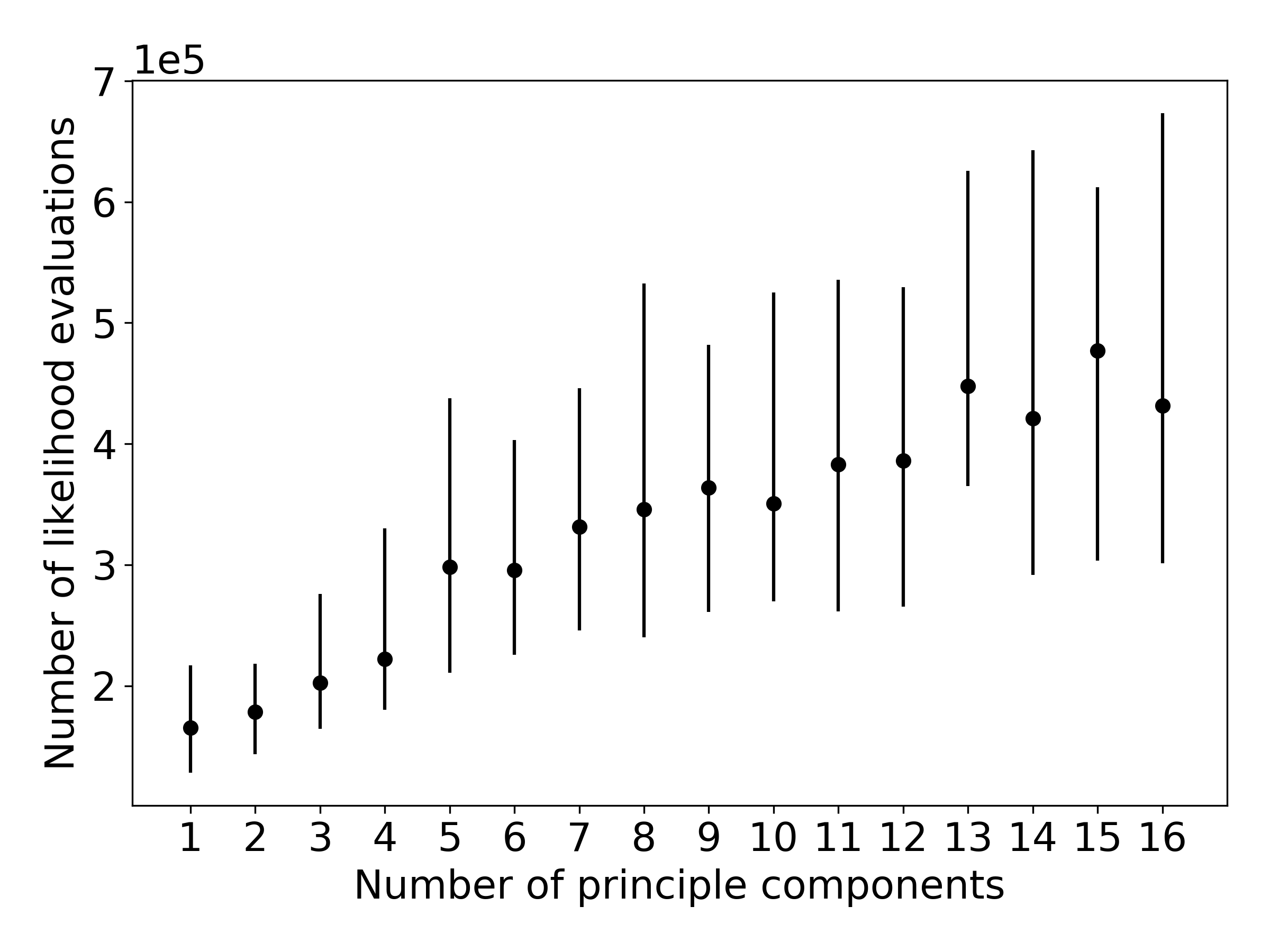}
    \caption{Number of likelihood evaluations needed to reach the nested sampling convergence criteria of $\Delta\log{Z}\leq0.3$, for an increasing number of included latent parameters. The error bars indicate the median and $68\%$ percent interval of the distributions for a sample of $50$ targets.}
    \label{fig:ncalls}
\end{figure}

Finally, we include a brief note on the run time of the sampler using different numbers of latent parameters. The nested sampling package \texttt{Dynesty} computes the model log-evidence, $\log{Z}$, during the sampling process, which decreases when sampling near the global likelihood maximum. As a stopping criterion we used the default value of $\Delta\log{Z} \leq 0.3$ for the change in the evidence over time, using the static sampler with a fixed value of $300$ live points. This allows us to estimate the change in the run time for different numbers of included latent parameters. The time to reach the stopping criterion is the product of the number of times the likelihood function is evaluated and the time required to evaluate it. The latter depends on the computational resources available, but also the length and cadence of the available flux time series. To evaluate the expected change in run time from changing the number of latent parameters, we therefore only considered the total number of likelihood evaluations needed to reach the stopping criterion, which is shown in Fig.~\ref{fig:ncalls}. While the absolute number of likelihood evaluations will change under different sampling schemes, the change in performance is likely conserved. From this comparison we find that time required to adequately sample the posterior distribution will likely improve by a factor of two to three, by a similar reduction in the number of latent variables used.   

\section{Discussion and conclusions} 
\label{sec:conclusions}
We have presented a method for reducing the complexity of sampling the model parameter space for the asteroseismic analysis of solar-like oscillators. The method uses \PCA to define a prior volume in a latent parameter space, which can be more easily sampled than the larger model parameter space. 

We tested the \PCA method using a spectrum model consisting of three Harvey-like background noise terms, a white noise term, and a series of Lorentzian peaks determined by the asymptotic relation for p modes. This model consists of $16$ parameters that are closely related to the physical properties of the star and four parameters that are likely instrument dependent and we therefore left as free variables. 

Identifying a single value of the number of latent parameters is likely not possible since our prior sample contains nonlinear correlations over wide ranges of $\numax$. Furthermore, the density of samples also changes dramatically at $\nu\approx244\muHz$ since the sample contains observations from both the long- and short-cadence modes of the \kepler mission as well as observations from the TESS mission. We therefore used the effective rank of the covariance matrix and the modified KG criterion to provide an indicator for the choice of the number of latent parameters. Based on these estimates, and the fact that stellar structure and oscillation codes suggest only approximately five dominant fundamental stellar parameters, our expectation is that the number of necessary latent parameters is likely between four and eight for the majority of this particular sample. 

From a comparison with reference frequencies and model parameters from \pbjam, we find that two latent parameters are, in general, sufficient to produce models that explain the bulk background features of the \PSD. This reflects the simple scaling relations for the background model parameters \citep[see, e.g.,][]{Kallinger2014}. Using more latent parameters, however, allows the model to more precisely explain the detailed features of the spectrum, such as the mode frequencies. We find that, using more than four latent parameters, the \PCA mode frequencies fall within $10\%$ of $\dotwo$ of the reference frequencies. Similarly, the majority of the asymptotic model parameters can be recovered to within a few percent by using more than two latent parameters. 

This suggests that, for the set of parameters in our prior sample, we are able to reduce the dimensionality of the parameter space by a factor of two to three, and thereby achieve a similar improvement in the time required to sample the posterior distribution. Going forward, this method can be used to add additional terms to the spectrum model, without necessarily increasing the computational complexity. For example, for main-sequence stars the $l=1$ and $l=3$ modes can likely be added without the need for additional latent parameters since their model parameters are highly correlated with those of the $l=0$ and $l=2$ modes. For stars evolving off the main sequence, this correlation becomes more complicated due to coupling with buoyancy-dominated modes in the stellar core. Additional latent parameters may therefore be necessary, but the total dimensionality of the sampling will likely still be less than the full model parameter space. This opens the potential for more complicated models to be more rapidly evaluated in a pipeline format, where stellar power density spectra can be analyzed on an ensemble scale. 

\begin{acknowledgements}
Thanks to Chris Moore for the useful and informative chats.

MBN acknowledges support from the UK Space Agency. 

GRD, and WJC acknowledge the support of the UK Science and Technology Facilities Council (STFC).

JO acknowledges support from NASA through the NASA Hubble Fellowship grant HST-HF2-51517.001 awarded by the Space Telescope Science Institute, which is operated by the Association of Universities for Research in Astronomy, Incorporated, under NASA contract NAS5-26555.

This paper has received funding from the European Research Council (ERC) under the European Union’s Horizon 2020 research and innovation programme (CartographY GA. 804752).

The authors acknowledge use of the Blue-BEAR HPC service at the University of Birmingham. 

This paper includes data collected by the \kepler mission and obtained from the MAST data archive at the Space Telescope Science Institute (STScI). Funding for the \kepler mission is provided by the NASA Science Mission Directorate. STScI is operated by the Association of Universities for Research in Astronomy, Inc., under NASA contract NAS5–26555. 

This paper includes data collected by the TESS mission. Funding for the TESS mission is provided by the NASA's Science Mission Directorate.

This work has made use of data from the European Space Agency (ESA) mission {\it Gaia} (\url{https://www.cosmos.esa.int/gaia}), processed by the {\it Gaia} Data Processing and Analysis Consortium (DPAC, \url{https://www.cosmos.esa.int/web/gaia/dpac/consortium}). Funding for the DPAC has been provided by national institutions, in particular the institutions participating in the {\it Gaia} Multilateral Agreement.

This publication makes use of data products from the Two Micron All Sky Survey, which is a joint project of the University of Massachusetts and the Infrared Processing and Analysis Center/California Institute of Technology, funded by the National Aeronautics and Space Administration and the National Science Foundation.

\end{acknowledgements}

\bibliographystyle{aa}
\bibliography{main, extra}

\appendix

\section{A numerical experiment with stellar models}
\label{app:model_var}
Here we consider a differential manifold $\mathcal{M}$, admitting an embedding $\varphi: \mathcal{M} \to \mathbb{R}^{N}$. By the Whitney embedding theorem, such an embedding is guaranteed to exist where $N$ is at least twice the intrinsic dimensionality of $\mathcal{M}$. In the neighborhood of a point $p \in \mathcal{M}$ described by coordinates $\{\theta_i: \mathcal{M} \to \mathbb{R}\}$ so that $\varphi(p) = \{x_j\}$, the push forward of the tangent vectors to $\mathcal{M}$ of these coordinates, $\left\{\partial_{\theta_i}\right\}$, have vector components in $\mathbb{R}^N$ that are simply the elements of the Jacobian matrix: $\varphi^* \partial_{\theta_j} = \frac{\partial x_i}{\partial \theta_j} \mathbf{e}_i$.

In reducing the dimensionality of the mode frequencies, we are implicitly supposing that the set of physically reasonable mode frequencies lie on just such a low-dimensional sub-manifold of what ought to otherwise be nominally a very high-dimensional space, in which any arbitrary set of mode frequencies represents a single point. From the above discussion, to explore the dimensionality of this physically reasonable sub-manifold, we may consider the Jacobian of the mode frequencies (which should collectively be the span of the coordinate basis vectors of the tangent space to this sub-manifold), with respect to a handful of parameters that a priori describe physical changes to the stellar structure (our physically motivated candidate latent variables). These parameters mathematically constitute candidate coordinates on this lower-dimensional manifold.

We computed these Jacobian matrix elements with respect to numerical models of stellar structure constructed using Modules for Experiments in Stellar Astrophysics
 \citep[MESA;][]{Paxton2011, Paxton2013, Paxton2015, Paxton2018, Paxton2019} r22.05.1, computing mode frequencies using GYRE \citep{Townsend2013}. We considered the following candidate latent variables:
(i) $R$, the stellar radius,
(ii) $M$, the stellar mass,
(iii) $Y_i$, the initial helium abundance,
(iv) $Z_i$, the initial metal fraction,
(v) $\alpha_\mathrm{MLT}$, the mixing-length efficiency parameter, and
(vi)  $f_\text{ov}$, which describes the scale length of MESA's implementation of exponential over-mixing.

In addition to these continuous variables, we also consider the following categorical variables:
(i) whether or not diffusion and settling of helium and heavy elements was simulated in constructing the stellar models,
(ii) which atmospheric boundary condition was applied (Eddington vs. Krishna-Swamy $T$-$\tau$ relations), and
(iii) which metal mixture was used (GS98 vs. AGS09).

We first performed this experiment for a solar-calibrated stellar model produced with MESA's simplex optimization tool, made with element diffusion, a small amount ($f_\mathrm{ov} = 0.01$) of overshoot, the Eddington $T-\tau$ relation, and the GS98 metal mixture. We computed $N$ mode frequencies at $l = 0, 1, 2$ within $\pm 5 \Delta\nu$ of $\nu_\mathrm{max}$. In the neighborhood of this reference model, we produce a set of perturbed models, where we perturb each of the continuous variables by 1\% relative to their nominal solar-calibrated values. From this perturbation, we are able to evaluate finite-difference derivatives of the mode frequencies with respect to each of these variables. For the categorical variables we only compute simple pairwise differences between identical modes.

We know a priori that the combination $M/R^3$ sets the fundamental frequency of p modes (via $\Delta\nu$), so rather than compute the Jacobian with respect to the dimensionful frequencies, we have used instead the Jacobian with respect to $\epsilon = \nu / \Delta\nu - n_p - l/2$. For each variable $i$ that we perturb, the dimensionless differences $\delta \epsilon_{i, n_p,l}$ form a row of the Jacobian matrix, of shape $9 \times N$. We then scale each row by its RMS value so that it yields a unit vector in a $N$-dimensional space: that is to say, we considered only the directions, and not the magnitudes, of the embedded candidate coordinate basis vectors. 
 
Having performed such rescaling, we then used \PCA in much the same fashion as we have the observational data. In particular, we considered the eigenvalues from singular value decomposition of the rescaled Jacobian matrix. Since the rescaled Jacobian matrix contains nine unit vectors, its squared eigenvalues should sum to 9. Accordingly, the fractional cumulative sum of squared eigenvalues (ordering the principal components in decreasing rank order) gives us the cumulative explained variance ratio in the mode frequencies induced by changes to this set of latent variables, just as in the case of the scree plot shown in \autoref{fig:skree} for \PCA on observational data.

\begin{figure}[htbp]
    \centering
    \includegraphics[width=.45\textwidth]{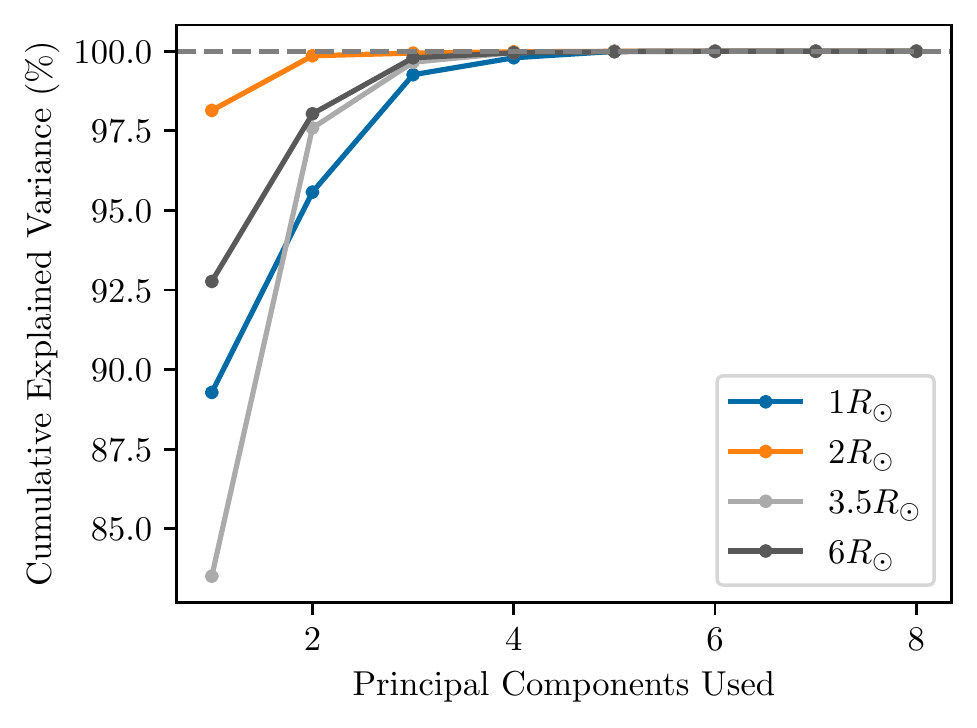}
    \caption{Cumulative explained variance ratios computed from the singular value decomposition of the Jacobian matrices for dimensionless $l = 0, 1, 2$ mode frequencies with respect to the nine candidate latent variables described in the main text. Different colors indicate values computed with respect to stellar models at fiducial radii of 1, 2, 3.5, and 6 $R_\odot$. In all other respects, these models are solar-calibrated. Since dimensionless frequencies are used in this calculation, and a combination of $M$ and $R$ sets the fundamental frequency for these normal modes via $\Delta\nu \sim \sqrt{M/R^3}$, one more degree of freedom must be considered than shown here for any given desired amount of variance to be explained.}
    \label{fig:mesa-var}
\end{figure}

We show this cumulative explained variance ratio for our solar-calibrated model with the blue points in \autoref{fig:mesa-var}, with $N = 30$. If the unit vectors that are its rows were all orthogonal, then all of these eigenvalues should be equal to 1, and so this ratio should increase linearly with the number of principal components used. However, we see instead that almost all of the variance can be explained with only two to three of these principal components (i.e., a total of three to four latent variables, once we also account for $\Delta\nu$). We repeat this procedure using as fiducial models a few later checkpoints along the solar-calibrated track, at radii of 2, 3.5, and 6 $R_\odot$. For these evolved stellar models, we restrict our attention to $l = 0, 2$ pure p-mode frequencies (and so have $N = 20$ in each case), evaluated by the $\pi$-mode prescription of \citet{Ong2020} as implemented in GYRE. These are also shown in \autoref{fig:mesa-var} using points and lines of different colors, and behave qualitatively very similarly to the main-sequence case. Thus, at least for pure p modes, only a handful of latent variables is required to describe the collective behavior of a very large number of mode frequencies, even at different stages of stellar evolution.

\end{document}